\documentclass[twocolumn,showpacs,aps,prd]{revtex4}

\usepackage{graphicx}
\usepackage{dcolumn}
\usepackage{amsmath}
\usepackage{epsfig}

\RequirePackage{xspace}

\newcommand{\GG}{\gamma\gamma}
\newcommand{\psip}{\psi^{\prime}}

\newcommand{\jpsi}{J/\psi}
\newcommand{\RDM}{{\cal R}_{Data/MC}}

\newcommand{\epl}{e^+}
\newcommand{\emi}{e^-}

\begin{document}

\def\fourpiz    {\ensuremath{\piz\piz\piz\piz}\xspace}
\def\psiprime   {\ensuremath{\psi^\prime}\xspace}
\def\fz#1       {\ensuremath{f_0({#1})}\xspace}
\def\Erad       {\ensuremath{E^*_\gamma}\xspace}
\def\besiii       {BESIII\xspace}
\def\stat       {_{\mathrm{stat}}}
\def\syst       {_{\mathrm{syst}}}
\def\etal       {{\em et al.}\xspace}
\def\ar         {\rightarrow}
\def\GG         {\gamma\gamma}
\def\bbt        {\bibitem}

\title{
{\boldmath Precision Measurement of the Mass of the $\tau$ Lepton} }
\date{\today}% It is always \today, today, but you may specify any date with \date.

\author{
\small
%\begin{small}
%\begin{center}
M.~Ablikim$^{1}$, M.~N.~Achasov$^{8,a}$, X.~C.~Ai$^{1}$, O.~Albayrak$^{4}$, M.~Albrecht$^{3}$, D.~J.~Ambrose$^{41}$, F.~F.~An$^{1}$, Q.~An$^{42}$, J.~Z.~Bai$^{1}$, R.~Baldini Ferroli$^{19A}$
% ...
\\
        \vspace{0.2cm}
        (BESIII Collaboration)\\
            \vspace{0.2cm} {\it
                $^{1}$ Institute of High Energy Physics, Beijing 100049, People's Republic of China\\
% ...
$^{3}$ Bochum Ruhr-University, D-44780 Bochum, Germany\\
                    $^{4}$ Carnegie Mellon University, Pittsburgh, Pennsylvania 15213, USA\\
% ...
$^{8}$ G.I. Budker Institute of Nuclear Physics SB RAS (BINP), Novosibirsk 630090, Russia\\
% ...
$^{19}$ (A)INFN Laboratori Nazionali di Frascati, I-00044, Frascati, Italy; (B)INFN and University of Perugia, I-06100, Perugia, Italy\\
% ...
$^{41}$ University of Rochester, Rochester, New York 14627, USA\\
                    $^{42}$ University of Science and Technology of China, Hefei 230026, People's Republic of China\\
% ...
$^{a}$ Also at the Novosibirsk State University, Novosibirsk, 630090, Russia\\
% ...
}
%\vspace{0.4cm}
%\end{small}
}

\newpage

\begin{abstract}

An energy scan near the $\tau$ pair production threshold has been
performed using the BESIII detector. About $24$ pb$^{-1}$ of data,
distributed over four scan points, was collected. This analysis is
based on $\tau$ pair decays to $ee$, $e\mu$, $eh$, $\mu\mu$, $\mu h$,
$hh$, $e\rho$, $\mu\rho$ and $\pi\rho$ final states, where $h$ denotes
a charged $\pi$ or $K$. The mass of the $\tau$ lepton is measured from
a maximum likelihood fit to the $\tau$ pair production cross
section data to be $m_{\tau} =
(1776.91\pm0.12~^{+0.10}_{-0.13}$)~MeV/$c^2$, which is currently the
most precise value in a single measurement.

\end{abstract}

% PACS, the Physics and Astronomy Classification Scheme.

\maketitle

\section{Introduction}

The $\tau$ lepton mass, $m_{\tau}$, is one of the fundamental
parameters of the Standard Model (SM). The relationship between the
$\tau$ lifetime ($\tau_{\tau}$), mass, its electronic branching
fraction ($B(\tau\rightarrow e\nu\bar{\nu})$) and weak coupling
constant $g_{\tau}$ is predicted by theory:
\begin{eqnarray}
 \frac{B(\tau\rightarrow e\nu\bar{\nu})}{\tau_{\tau}}=\frac{g_{\tau}^2
 m_{\tau}^5}{192\pi^3},
\label{Eq.calgtau}
\end{eqnarray}
up to small radiative and electroweak
corrections~\cite{LepUnivCor}. It appeared to be badly violated before
the first precise $m_{\tau}$ measurement of BES became available in
1992~\cite{bes1}; this measurement was later updated with more $\tau$
decay channels~\cite{bes2prd} and confirmed by subsequent measurements
from BELLE~\cite{belle}, KEDR~\cite{kedr}, and BABAR~\cite{BABAR}. The
experimental determination of $\tau_{\tau}$, $B(\tau\rightarrow
e\nu\bar{\nu})$ and $m_{\tau}$ to the highest possible precision is
essential for a high precision test of the SM. Currently, the mass
precision for $e$ and $\mu$ has reached $\Delta m/m$ of
$10^{-8}$ , while for $\tau$ it is $10^{-4}$~\cite{pdg}.

A precision $m_{\tau}$ measurement is also required to check lepton
universality. Lepton universality, a basic ingredient in the minimal
standard model, requires that the charged-current gauge coupling
strengths $g_{e}$, $g_{\mu}$, $g_{\tau}$ should be identical:
$g_e=g_{\mu}=g_{\tau}$. Comparing the electronic branching fractions
of $\tau$ and $\mu$, lepton universality can be tested as:
\begin{eqnarray}
    \left(\frac{g_{\tau}}{g_{\mu}}\right)^{2}=\frac{\tau_{\mu}}{\tau_{\tau}}\left(\frac{m_{\mu}}{m_{\tau}}\right)^{5}\frac{B(\tau\rightarrow
            e\nu\bar{\nu})}{B(\mu\rightarrow
                e\nu\bar{\nu})}(1+F_{W})(1+F_{\gamma}),
\label{Eq.universaltest}
\end{eqnarray}
where $F_{W}$ and $F_{\gamma}$ are the weak and
electromagnetic radiative corrections~\cite{LepUnivCor}.  Note
$(g_{\tau}/g_{\mu})^2$ depends on $m_{\tau}$ to the fifth power.

Furthermore, the precision of $m_{\tau}$ will also restrict the
ultimate sensitivity of $m_{\nu_{\tau}}$. The most sensitive bounds on
the mass of the $\nu_{\tau}$ can be derived from the analysis of the
invariant-mass spectrum of semi-hadronic $\tau$ decays, e.g. the
present best limit of $m_{\nu_{\tau}}<18.2$~MeV/$c^2$ ($95\%$
confidence level) was based on the kinematics of $2939$ ($52$)
events of $\tau^{-}\rightarrow 2\pi^{-}\pi^{+}\nu_{\tau}$
($\tau^{-}\rightarrow 3\pi^{-}2\pi^{+}(\pi^{-})\nu_{\tau}$)
~\cite{mntau}. This method depends on a determination of the
kinematic end point of the mass spectrum; thus high precision on
$m_{\tau}$ is needed.

So far, the pseudomass technique and the threshold
scan method have been used to determine $m_{\tau}$. The former, which
was used by ARGUS~\cite{argus}, OPAL~\cite{opal}, BELLE~\cite{belle}
and BABAR~\cite{BABAR}, relies
on the reconstruction of the invariant-mass and energy of the hadronic
system in the hadronic $\tau$ decay, while the latter, which was used
in DELCO~\cite{delco}, BES~\cite{bes1, bes2prd} and KEDR~\cite{kedr}, is a
study of the threshold behavior of the $\tau$ pair production cross
section in $e^{+}e^{-}$ collisions and it is the method used in this
paper. Extremely important in this approach is to determine the beam
energy and the beam energy spread precisely.  Here the beam energy
measurement system (BEMS)~\cite{bems} for BEPCII is used and will be
described below.

Before the experiment began, a study was carried out
using Monte Carlo (MC) simulation and sampling to optimize the
number and choice of scan points in order to provide the highest
precision on $m_{\tau}$ for a specified period of data taking time
or equivalently for a given integrated luminosity~\cite{optstd3}.

The $\tau$ scan experiment was done in December 2011. The $J/\psi$ and
$\psi^{\prime}$ resonances were each scanned at seven energy
points, and data were collected at four scan points near $\tau$ pair
production threshold with center of mass (CM) energies of 3542.4~MeV, 3553.8~MeV,
3561.1~MeV and 3600.2~MeV.  The first $\tau$ scan point is below the
mass of $\tau$ pair~\cite{pdg}, while the other three are
above.

%%%%%%%%%%%%%%%%%%%%%%%%%%%%%%%%%%%%%%%%%%%%%%%%%%%%%%%%%%%%%%%%%%%%%%%%%%%
\section{BESIII Detector}

The BESIII detector is designed to study hadron spectroscopy and
$\tau$-charm physics~\cite{bes3}. The cylindrical BESIII is composed
of: (1) A Helium-gas based Main Drift Chamber (MDC) with 43 layers
providing an average single-hit resolution of 135~$\mu$m, and
a charged-particle momentum resolution in a 1 T magnetic field of 0.5\%
at 1.0~GeV/$c$. (2) A Time-of-Flight (TOF) system constructed of
5-cm-thick plastic scintillators, with 176 counters of 2.4~m length in
two layers in the barrel and 96 fan-shaped counters in the end-caps.
The barrel (end-cap) time resolution of 80 ps (110 ps) provides
2$\sigma$ $K/\pi$ separation for momenta up to 1.0~GeV/$c$. (3) A
CsI(Tl) Electro-Magnetic Calorimeter (EMC) consisting of 6240 crystals
in a cylindrical barrel structure and two end-caps. The energy
resolution at 1.0~GeV/$c$ is 2.5\% (5\%) in the barrel (end-caps), while
the position resolution is 6 mm (9 mm) in the barrel (end-caps). (4)
A Resistive plate chamber (RPC)-based muon chamber (MUC) consisting of
1000~m$^2$ of RPCs in nine barrel and eight end-cap layers and
providing 2~cm position resolution.
%The photon energy resolution at BESIII is much better than that at BESII
%and comparable to those at CLEO~\cite{cleod} and Crystal
%Ball~\cite{cball}.

%%%%%%%%%%%%%%%%%%%%%%%%%%%%%%%%%%%%%%%%%%%%%%%%%%%%%%%%%%%%%%%%%%%%%%%%%%%
\section{Beam Energy Measurement System}\label{sect_bsbs}
\subsection{Introduction}

\begin{figure*}
\includegraphics{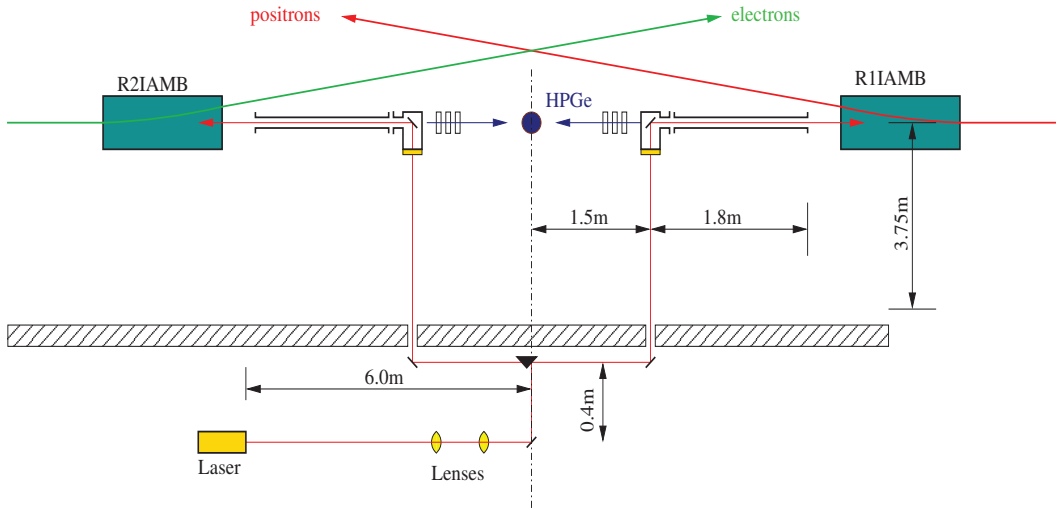}% Here is how to import EPS art
\caption{\label{fig:bemsys}Simplified schematic of the beam energy
measurement system. The positron and electron beams are indicated.
R1IAMB and R2IAMB are accelerator magnets, and the HPGe detector is
represented by the dot at the center. The half-meter shielding wall of
the beam tunnel is shown cross-hatched.  The laser and optics system
is located outside the tunnel of the storage ring, where the optics
system are composed of two lenses, mirrors and a prism
denoted by the inverted solid triangle.}
\end{figure*}

The BEMS is located at the north crossing point of the BEPCII storage
ring.  The layout schematic of BEMS is shown in Fig.~\ref{fig:bemsys}.
This design allows us to measure the energies of both the electron and
positron beams with one laser and one High Purity Germanium (HPGe)
detector~\cite{bems}.

In the Compton scattering process, the maximal energy of the
scattered photon $E_{\gamma}$ is related to the electron energy
$E_e$ by the kinematics of Compton scattering~\cite{Rullhusen,Landau}:
\begin{equation}
E_e = \frac{E_{\gamma}}{2} \left[ 1 + \sqrt{1+\frac{m_e^2}{{\cal E_{\gamma}} E_{\gamma}} } \right],
\label{Eq.energy}
\end{equation}
where ${\cal E_{\gamma}}$ is the energy of the initial photon, i.e. the
energy of the laser beam in the BEMS. The scattered photon energy can be
measured with high accuracy by the HPGe detector, whose energy scale
is calibrated with photons from radioactive sources and the readout
linearity is checked with a precision pulser. The maximum
energy can be determined from the fitting to the edge of the
scattered photon energy spectrum.  At the same time the energy spread
of back-scattered photons due to the energy distribution of the collider
beam is obtained from the fitting procedure~\cite{bems}. Finally, the electron
energy can be calculated by Eq.~\ref{Eq.energy}. Since the energy of
the laser beam and the electron mass ($m_e$) are determined with the
accuracy at the level of $10^{-8}$, $E_e$ can be determined, utilizing
Eq.~\ref{Eq.energy}, as accurately as $E_{\gamma}$, an accuracy
is at the level of $10^{-5}$.
The systematic error of the electron and positron beam
energy determination in our experiment was tested
through previous measurement of the $\psi^{\prime}$ mass and was estimated as
$2\times 10^{-5}$~\cite{bems}; the relative uncertainty of the beam
energy spread was about $6\%$~\cite{bems}.

\subsection{Determination of Scan Point Energy}
The BEMS alternates between measuring electron and positron beam
energies, and writes out energy calibration (EC) data files. Each EC
file has its own time stamp that can be used to associate BEMS
measurements with corresponding scan data. In the $\tau$-scan region,
all EC runs within the start and end times of a scan point are used
for determining the scan-point energy.

In the case of fast energy scans in the $\jpsi$ and $\psip$ resonance
regions, the ratio of hadronic and Bhabha events is used to determine
scan-point boundaries.  Once all electron and positron EC files that
belong to a particular scan point are grouped, we determine the CM
energy of the crossing beams at the given scan-point by using the
error-weighted average of electron and positron beam energies,
$\overline{E}_{e^-}$ and $\overline{E}_{e^+}$, respectively.  The CM
energy of a scan-point is calculated using
\begin{equation}\label{eq:ecm}
E_{\rm{CM}} = 2 \sqrt{ \overline{E}_{e^-} \cdot \overline{E}_{e^+} } \cdot \cos{(\theta_{e^+e^-}/2)},
\end{equation}
where $\theta_{e^+e^-} = 0.022$~rad is the crossing angle between the
beams.  Table~\ref{lum} gives measured luminosities ($\mathcal{L}$) at
each scan-point, in which the CM energy is obtained from Eq.~\ref{eq:ecm};
the method to determine these luminosities will be introduced in
Section~\ref{sec-lum}.

\begin{table}[tb]
\begin{center}
  \caption{Measured integrated luminosities at each scan-point. The
  errors are statistical only.}
\begin{tabular} {lccc}
\hline \hline
Scan   &  $E_{\rm{CM}}$ (MeV)  & $\mathcal{L}$(nb$^{-1}$) \\
\hline
$J/\psi$ & 3088.7  & $78.5 \pm 1.9$  \\
          & 3095.3 & $219.3 \pm 3.1$ \\
          & 3096.7 & $243.1 \pm 3.3$ \\
          & 3097.6 & $206.5 \pm 3.1$ \\
          & 3098.3 & $223.5 \pm 3.2$ \\
          & 3098.8 & $216.9 \pm 3.1$ \\
          & 3103.9 & $317.3 \pm 3.8$ \\
% \\
$\tau$ & 3542.4 & $4252.1 \pm 18.9$\\
       & 3553.8 & $5566.7 \pm 22.8$\\
       & 3561.1 & $3889.2 \pm 17.9$\\
       & 3600.2 & $9553.0 \pm 33.8$\\
% \\
$\psi'$ & 3675.9 & $787.0  \pm 7.2$   \\
        & 3683.7 & $823.1  \pm 7.4$   \\
        & 3685.1 & $832.4  \pm 7.5$   \\
        & 3686.3 & $1184.3 \pm 9.1$   \\
        & 3687.6 & $1660.7 \pm 11.0$  \\
        & 3688.8 & $767.7  \pm 7.2$   \\
        & 3693.5 & $1470.8 \pm 10.3$  \\
\hline\hline
\end{tabular}
\label{lum}
\end{center}
\end{table}

\subsection{Determination of Beam Energy Spread}
Besides measuring the energy of the electron or positron beams, the
BEMS also measures the energy spread independently of the energy
measurements by the accelerator. Using the same grouping of the EC data,
we obtain weighted averages of the electron, $\delta_{e^-}$, and the positron, 
$\delta_{e^+}$, energy spreads. Corresponding errors, $\Delta(\delta_{e^-})$ and
$\Delta(\delta_{e^+})$, represent one standard deviation of
weighted-averages.  Taking into account that the beam energy has a
Gaussian distribution around its mean with the width given by the
energy spread, the total energy spread of a scan point, $\delta_{w}^{\rm{BEMS}}$,
is calculated from the average electron and positron spreads using:
\begin{equation}\label{eq:spread}
\delta_{w}^{\rm{BEMS}} = \sqrt{ \delta^2_{e^-} + \delta^2_{e^+} }.
\end{equation}
\begin{figure*}
\includegraphics[height=4.0cm]{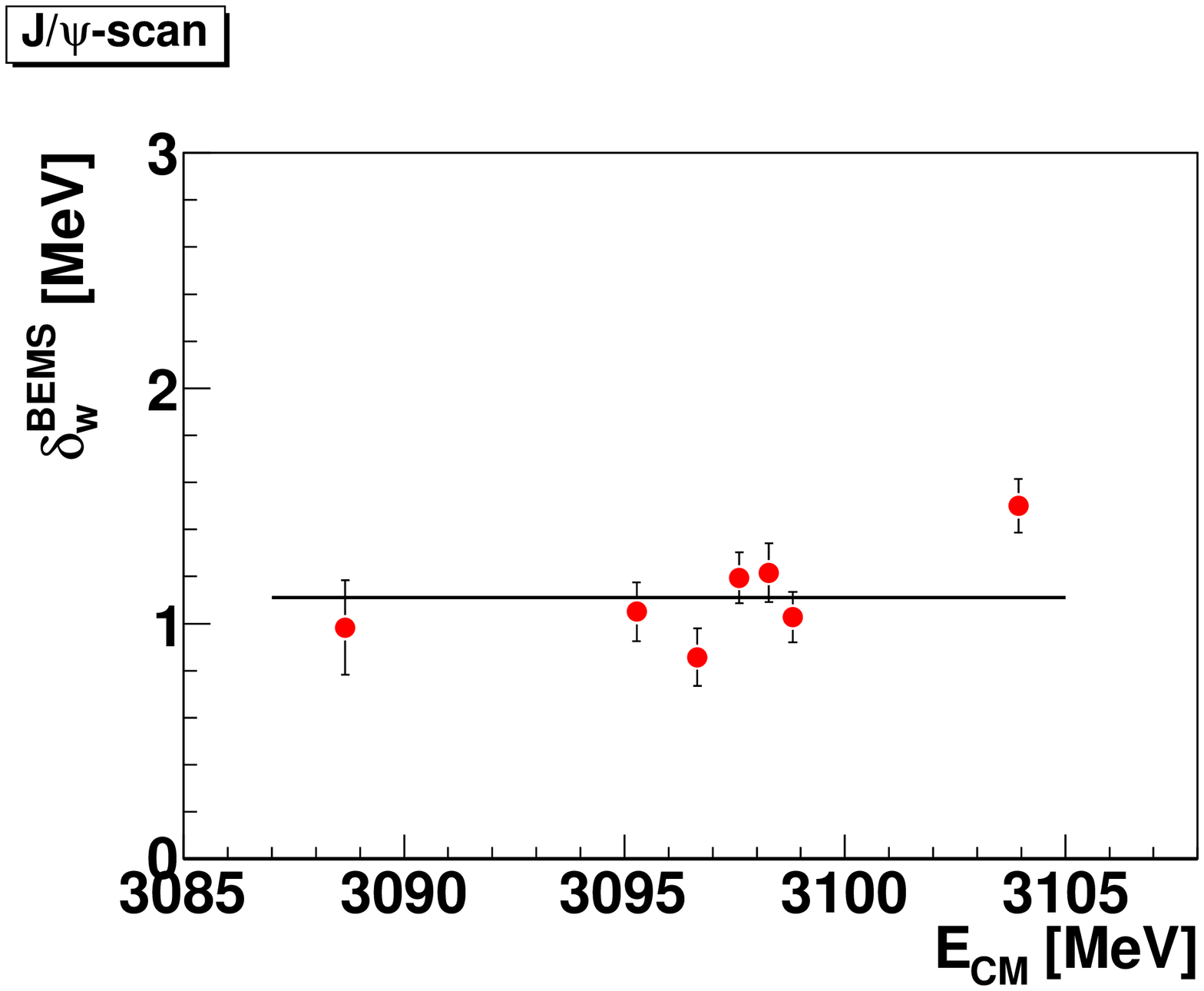}
\includegraphics[height=4.0cm]{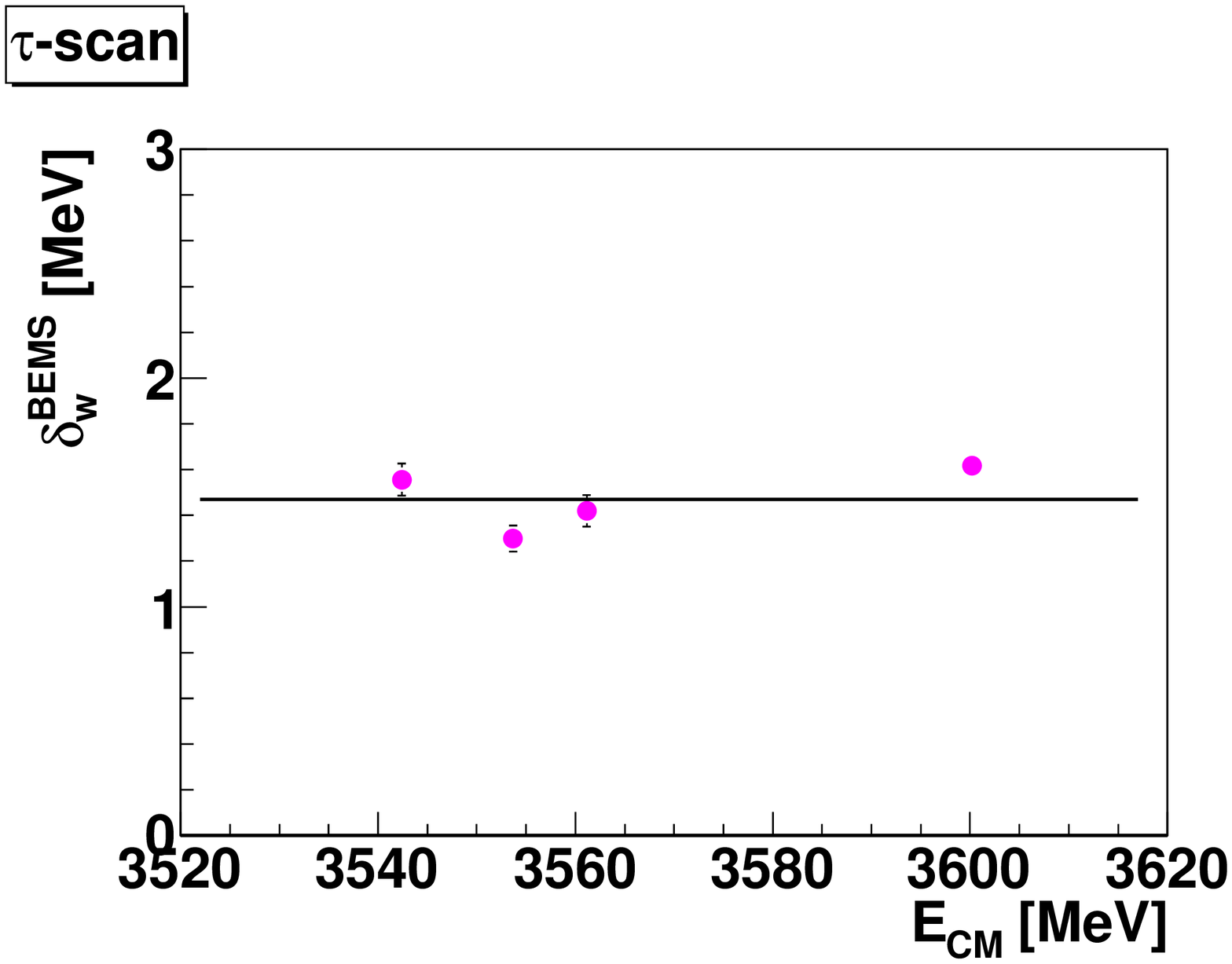}
\includegraphics[height=4.0cm]{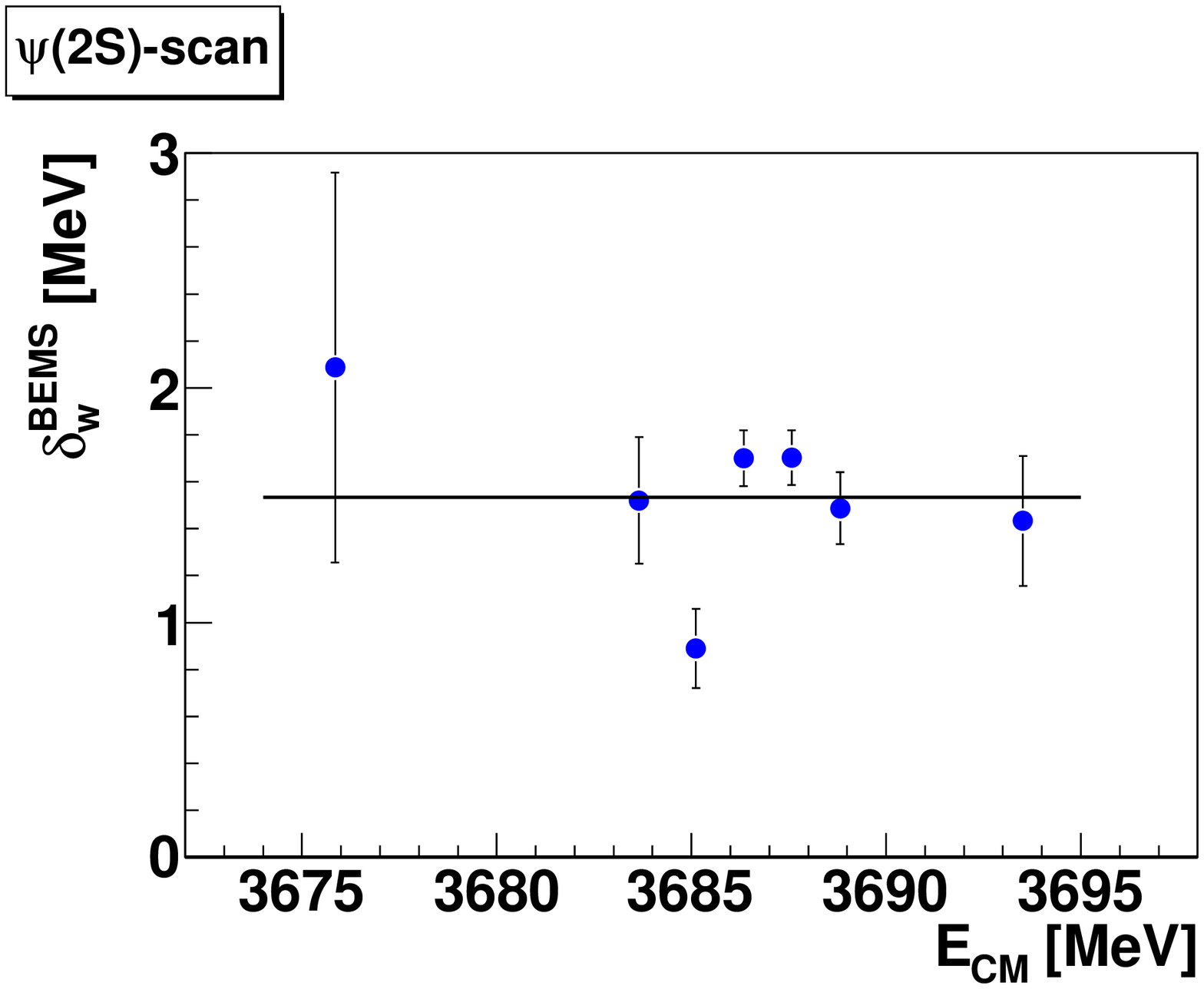}
\caption{\label{fig:wSpread}Energy spreads from the $\jpsi$
(left), $\tau$ (middle) and $\psip$ scans (right).  Horizontal lines
represent mean values, listed in Table~\ref{tab:wSpreadSP}.}
\end{figure*}
It is assumed that the $e^-$ and $e^+$ EC measurements are independent
and that the total beam energy spread results from the sum of two
uncorrelated Gaussian distributions.  The crossing angle between the
beams has little effect on the energy spread and is ignored.
Consequently, the error on the total spread is obtained using error
propagation:
\begin{equation}\label{eq:spreadErr}
\Delta(\delta_{w}^{\rm{BEMS}}) = \sqrt{ \delta^2_{e_-} \cdot \Delta^2(\delta_{e_-}) + \delta^2_{e_+} \cdot \Delta^2(\delta_{e_+})  } 
                   / \delta_w^{\rm{BEMS}}.
\end{equation}

Figure~\ref{fig:wSpread} shows the corresponding energy spreads from the
$\jpsi$ (left), $\tau$ (middle) and $\psip$ (right) scan regions.  The
spreads show little dependence on energy within a given scan region,
and increase gradually as the CM energy increases from the $\jpsi$ to
the $\psip$ region.  Because of the large fluctuations, the energy
spread in each scan region is estimated by calculating the mean energy
spread, taking the error as one standard deviation of the mean. The mean
values are summarized in Table~\ref{tab:wSpreadSP}, and shown as
horizontal lines on the plots in Fig.~\ref{fig:wSpread}.

\begin{table}[tb]
\begin{center}
  \caption{Energy spreads (MeV) from the
$\jpsi$, $\tau$ and $\psip$ scan regions
calculated using Eqs.~(\ref{eq:spread}) and~(\ref{eq:spreadErr}).}
\begin{tabular} {lccc}
\hline \hline
Scan   & $\delta_w^{\rm{BEMS}}$ & $\Delta(\delta_w^{\rm{BEMS}})$ \\
\hline
$\jpsi$ & 1.112  &  0.070  \\
$\tau$  & 1.469  &  0.064  \\
$\psip$ & 1.534  &  0.109 \\
\hline\hline
\end{tabular}
\label{tab:wSpreadSP}
\end{center}
\end{table}

%%%%%%%%%%%%%%%%%%%%%%%%%%%%%%%%%%%%%%%%%%%%%%%%%%%%%%%%%%%%%%%%%%%%%%%%%%%
\section{The Data Sample and MC Simulation}
The $J/\psi$ and $\psi^{\prime}$ scan data samples listed in
Table~\ref{lum} are used to determine the line shape of each
resonance, and the parameters obtained are used to validate the BEMS
measurements. All of the data collected near $\tau$ pair production
threshold are used to do the $\tau$ mass measurement.

The luminosity at each scan point is determined using two-gamma events
($e^+ + e^- \to \gamma \gamma (\gamma))$. Bhabha events are used to do
a cross check. The Babayaga 3.5 generator~\cite{babayaga} is used as
our primary generator.

% MC samples used in J/psi and psi' line shaple need to be added.

To devise selection criteria for hadronic events in resonance scans,
we analyzed $\approx 10^{6}$ events from the $\jpsi$ and $\psip$ data
and $\approx 50 \times 10^6$ events from the continuum data produced
at center-of-mass energies 3096~MeV, 3686~MeV and 3650~MeV,
respectively. Approximately $5 \times 10^6$ events from corresponding
$\jpsi$ and $\psip$ inclusive MC samples are also used to optimize the
selection criteria.
%we analysed the $\jpsi$, $\psip$ and continuum data produced at
%center-of-mass energies 3.096 GeV, 3.686 GeV and 3.65 GeV,
%respectively. Corresponding $\jpsi$ and $\psip$ MC samples
%are also used for the signal/background assessment.

The {\sc GEANT4}-based~\cite{geant4} simulation software, BESIII Object
Oriented Simulation~\cite{boss}, contains the detector geometry and
material description, the detector response and signal digitization
models, as well as the detector running conditions and performance.
The production of the $\jpsi$ and $\psi^\prime$ resonance is simulated
by the Monte Carlo event generator {\sc kkmc}~\cite{kkmc}; the known
decay modes are generated by {\sc evtgen}~\cite{evtgen} with branching
ratios set at Particle Data Group (PDG)~\cite{pdg} world average
values, and by {\sc lundcharm}~\cite{lund} for the remaining unknown
decays.

{\sc kkmc}~\cite{kkmc} is also used to simulate the production of
$\tau$ pairs, and {\sc evtgen}~\cite{evtgen} is used to generate all
$\tau$ decay modes with branching ratios set at PDG~\cite{pdg} world
average values.  The MC sample including all of possible decay
channels is used as the $\tau$ pair inclusive MC sample; while the sample
including only a specific decay channel is used as a $\tau$ exclusive
MC sample.

%%%%%%%%%%%%%%%%%%%%%%%%%%%%%%%%%%%%%%%%%%%%%%%%%%%%%%%%%%%%%%%%%%%%%%%%%%%

\section{Event Selection Criteria}
Four data samples, including two-gamma events, Bhabha events, hadronic
events and $\tau$ pair candidate events, are used in this analysis.
Selection criteria to select these samples with high efficiency while
removing background are listed below.

\subsection{Good Photon-selection Criteria}
A neutral cluster is considered to be a good photon candidate if the
deposited energy is larger than 25~MeV in the barrel EMC
($|\cos\theta|<0.8$) or 50~MeV in the end-cap EMC
($0.86<|\cos\theta|<0.92$), where $\theta$ is the polar angle of the
shower.

\subsection{Good Charged Track Selection Criteria}
Good charged tracks are required to satisfy $V_{r}=\sqrt{V_x^2+V_y^2}<1$
cm, $|V_z|<10$ cm. Here $V_x$, $V_y$ and $V_z$ are the $x$, $y$ and
$z$ coordinates of the point of closest approach to the interaction
point (IP), respectively. The track is also required to lie
within the region $|\cos\theta|<0.93$.

\subsection{Two-Gamma Events}

The number of good photons is required to be larger than one and less
than eleven; the energy of the highest energy photon must be larger
than 0.85$\times E_{\rm{CM}}$/2 and less than 1.1$\times E_{\rm{CM}}$/2; the
energy of the second highest energy photon must be larger than 0.57
$\times E_{\rm{CM}}$/2; and the difference of the azimuthal angles of the
two highest energy photons in the CM must satisfy: $176^{\circ} <
\Delta \phi < 183^{\circ}$.  It is also required that there are no
charged tracks in the events.

\subsection{Bhabha Events}

The charged tracks must satisfy: $V_{r}<2$~cm, $|V_z|<10$~cm,
$|\cos\theta|<0.80$ and have momenta $p<2500$~MeV$/c$.  A good photon
must have deposited energy in the EMC less than 1.1$\times E_{\rm{CM}}$/2
and have $0<t<750$~ns, where $t$ is the time information from the EMC, to
suppress electronic noise and energy deposits unrelated to the event.
For the whole event, the following selection criteria are used: the
visible energy of the event must be larger than 0.22$\times E_{\rm{CM}}$;
the number of charged tracks is required to be two or three; the
momentum of the highest momentum charged track must be larger than
0.65$\times E_{\rm{CM}}$/2; the ratio $E/cp$ of one of the two highest
momentum tracks must be larger than 0.6, where $E$ is the energy deposited in
the calorimeter and $p$ is the track momentum determined by the MDC;
and the difference of the azimuthal angles of the two high momentum
tracks in the CM system must satisfy: $175^{\circ} < \Delta \phi <
185^{\circ}$.

\subsection{Hadronic Events}
Aside from standard selection requirements for good charged tracks,
the average vertex position along the beam line is required to
satisfy $ |\overline{ V_{Z} }| = |\frac{\sum_{i}^{N_{ch}} V_z^i
}{N_{ch}}| < 4$~cm, and the number of charged tracks ($N_{ch}$) must be larger
than two.

\subsection{\boldmath $\tau$ Pair Candidate Events}

In order to reduce the statistical error in the $\tau$ lepton mass,
this analysis incorporates 13 two-prong $\tau$ pair final states,
which are $ee$, $e\mu$, $e\pi$, $eK$, $\mu\mu$, $\mu\pi$, $\mu K$,
$\pi K$, $\pi\pi$, $KK$, $e\rho$, $\mu\rho$ and $\pi\rho$, with
accompanying neutrinos implied.  For the first ten decay channels,
there is no photon; for $X\rho$ ($X = e$, $\mu$ or $\pi$), the $\rho$
candidate is reconstructed with $\pi^{\pm}\pi^{0}$, so there are two
photons in the final state. No photons are allowed except in the
$\rho$ case where only two are allowed.  The number of good charged
tracks and also the number of total charged tracks are required to be
two for all channels.  The following event selection criteria are
applied to both data and MC samples.

\subsubsection{Additional Requirements on Good Photons}
Apart from those basic requirements, good photons must have the angle
between the cluster and the nearest charged particle larger than 20
degrees. Also we require $0<t<750$ ns.%, here t is the
%time information from the EMC to suppress electronic noise and
%energy deposits unrelated to the event.

\subsubsection{PID for Each Charged Track} 
For each charged track, the measured $p$, $E$, $E/cp$,
the time-of-flight value, the depth of the track in
the MUC ($D$) and the total number of hits in the MUC ($N_h$) are used
together to identify the particle type; the particle identification
(PID) criteria are listed in Table~\ref{pid}. In this table, $\Delta
TOF(e)$ is the difference between the calculated time-of-flight of the
track when it is assigned as an electron and the time-of-flight
measured by TOF; $\Delta TOF(\mu)$, $\Delta TOF(\pi)$ and $\Delta
TOF(K)$ are similar quantities. $p_{min}$ and $p_{max}$ are the
minimum and maximum momentum of charged tracks in any $\tau$ decay at
a given CM energy, which are all determined from the signal MC
simulation and are different in different scan energy points as
$p$ of these daughter particles are related with the initial
momentum of $\tau^{\pm}$. For $\pi^{\pm}$ from $\rho^{\pm}$, the
$p$ requirement is removed for the PID.

\begin{table*}[tb]
\begin{center}
  \caption{PID for charged particles. For the first scan point, the
  values of $p_{min}$ ($p_{max}$) are 0.2 GeV/$c$ (0.92 GeV/$c$), 0.2
  GeV/$c$ (0.9 GeV/$c$), 0.84 GeV/$c$ (0.93 GeV/$c$), and 0.76 GeV/$c$
  (0.88 GeV/$c$) for $e$, $\mu$, $\pi$, and $K$, respectively.}
\begin{tabular} {lcccccc}
\hline \hline
PID & $p$ (MeV$/c$) & EMC & TOF & MUC & other\\
\hline
  $e$ &$p_{min}<p<p_{max}$ & 0.8 $<E/cp<$ 1.05 &$|\Delta TOF(e)|<$0.2 ns &  &  \\
      &&&0 ns$<TOF<$4.5 ns&&\\  \hline
  $\mu$ & $p_{min}<p<p_{max}$ & $E/cp<$ 0.7 &$|\Delta TOF(\mu)|<$0.2 ns&($D>$(80$\times p$-50) cm or $D>$40 cm)&\\
  &&0.1$<E<$0.3&&and $N_h>$1&\\ \hline
  $\pi$ & $p_{min}<p<p_{max}$ & $E/cp<$ 0.6 &$|\Delta TOF(\pi)|<$0.2 ns&&not $\mu$\\
  &&&0 ns$<TOF<$4.5 ns&& \\ \hline
  $K$ & $p_{min}<p<p_{max}$ & $E/cp<$ 0.6 &$|\Delta TOF(K)|<$0.2 ns&&not $\mu$\\
  &&&0 ns$<TOF<$4.5 ns&& \\
\hline\hline
\end{tabular}
\label{pid}
\end{center}
\end{table*}

\subsubsection{Other Additional Requirements}
\label{subsubsec-selcut}

For the $X\rho$ channels, the invariant-mass of the two photons
($M(\gamma\gamma)$) is required to be in the $\pi^{0}$ mass window
which is [112.8, 146.4]~MeV/$c^2$.  Then these two photons are used
together with a $\pi$ candidate to reconstruct a $\rho$ candidate, and
the invariant-mass of the $\rho$ candidate is required to be in the
mass window i.e. [376.5, 1195.5]~MeV/$c^2$. Also, the magnitude of
the momentum of the $\rho$ candidate must be more than the minimum
expected momentum ($p_{min}^{\rho}$) and less than the maximum
($p_{max}^{\rho}$), where $p_{min}^{\rho}$ and $p_{max}^{\rho}$ are
also determined from the $p$ distribution of $\rho$ candidates in
the signal MC samples.

\begin{figure}
\includegraphics[width=0.7\columnwidth]{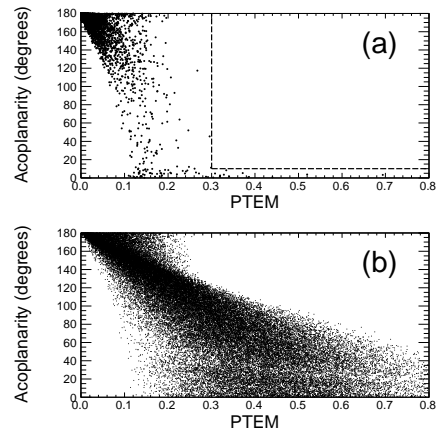}
\caption{\label{eeacopvsptem}(a) The scatter plot of PTEM versus
acoplanarity for the $\tau$ pair candidate $ee$ event from the first
scan energy point, which is below $\tau$ pair production threshold. 
The region above and to the right of the dashed line are the acceptance region.
(b) The same scatter plot for $ee$ events obtained from
the $\tau$ pair MC simulation corresponding to the second scan point,
which is above $\tau$ pair production threshold, after applying
the same selection criteria as used for data.}
\end{figure}

The $\tau$ pair candidate $ee$ event sample contains background from
two-photon $e^+e^-\rightarrow e^+(e^-e^+)e^-$ events in which the
leading $e^+$ and $e^-$ in the final state are undetected. These QED
background events are characterized by small net observed transverse
momentum and large missing energy. It follows that the variable PTEM,
defined as
\begin{eqnarray}
\label{ptem} PTEM=\frac{P_{T}}{E^{max}_{miss}}=\frac{(c\vec P_1+c\vec
P_2)_T}{W-|c \vec P_1|-|c \vec P_2|},
\end{eqnarray}
which is the ratio of the net observed transverse momentum to the
maximum possible value of the missing energy, is localized to small
values for QED background events. The first point in the $\tau$ mass
scan experiment is located below the $\tau$ pair production threshold
(about $11.2$~MeV below the mass of the $\tau$ pair, where the $\tau$
mass from the PDG is used), so all events passing the criteria for
selecting $\tau$ pair candidates at this point are background, and can
be used to study the event selection criteria and the background level
at the same time. The correlation between PTEM and the acoplanarity
angle $\theta_{acop}$ is studied for the background data set and the
signal MC sample. The acoplanarity angle $\theta_{acop}$ is defined as
the angle between the planes spanned by the beam direction and the
momentum vectors of the two final state charged tracks; i.e., it is
the angle between the transverse momentum vectors of the two
final state charged tracks.  Figures~\ref{eeacopvsptem}(a)
and ~\ref{eeacopvsptem}(b) are the distributions of PTEM versus $\theta_{acop}$ for $ee$ candidate
events from the first scan energy point data set, and $ee$ events from
the $\tau$ pair MC sample corresponding to the second scan point,
respectively.

From the comparison of these two plots, we retain only those
$ee$ events having PTEM$ > $0.3 and $\theta_{acop}> 10^{\circ}$.
By comparing the scatter plots of PTEM versus
$\theta_{acop}$ from the first scan point data set and that from the
second scan point signal MC simulation sample, we obtain similar
requirements for the other $\tau$ pair decay channels, which are
listed in Table~\ref{addcuts}.

\begin{table}[tb]
\begin{center}
  \caption{Selection requirements on acoplanarity
angle and PTEM for different final states.}
\begin{tabular} {lccc}
\hline \hline
  final state & $\theta_{acop}$ & PTEM\\
\hline
  $ee$ &  $>$10$^{\circ}$ & $>$0.3\\
  $e\mu$ &  $<$160$^{\circ}$ & $>$0.1\\
  $e\pi$ &  $<$170$^{\circ}$ & $>$0.1\\
  $eK$ &  $<$170$^{\circ}$ & \\
  $\mu\mu$ &  $<$140$^{\circ}$ & \\
  $\mu h$ &  $<$140$^{\circ}$ & \\
  $hh$ &$<$160$^{\circ}$& \\
  $e\rho$ &$<$170$^{\circ}$& \\
  $\mu\rho$ &$<$150$^{\circ}$& \\
  $\pi\rho$ &  & \\
\hline\hline
\end{tabular}
\label{addcuts}
\end{center}
\end{table}

%%%%%%%%%%%%%%%%%%%%%%%%%%%%%%%%%%%%%%%%%%%%%%%%%%%%%%%%%%%%%%%%%%%%%%%%%%%

\section{Data Analysis}
\subsection{Luminosity at Each Scan Point}
\label{sec-lum}

For all scan points, the luminosity $\mathcal{L}$ is determined from
$\mathcal{L}$ = $N_{data}/\epsilon_{\GG} \sigma_{\GG}$, where $N_{data}$ is the
number of selected two-gamma events in data and $\epsilon_{\GG}$ and
$\sigma_{\GG}$ are the efficiency and the cross section determined by the
Babayaga 3.5 MC, respectively. The measured luminosity ($\mathcal{L}$)
at each scan-point is listed in Table~\ref{lum}, from which the
integrated luminosities for the $J/\psi$, $\tau$ and $\psi^{\prime}$
scan are calculated as 1505~nb$^{-1}$, 23261~nb$^{-1}$ and
$7526$~nb$^{-1}$, respectively.  The analysis using Bhabha events is done
as a cross check of the two-gamma luminosity and gives consistent
luminosity results within 2\%.  The Bhabha luminosities
will also be used in the systematic error analyses.

%%%%%%%%%%%%%%%%%%%%%%%%%%%%%%%%%%%%%%%%%%%%%%%%%%%%%%%%%%%%%%%%%%%%%%%%%%%
\subsection{\boldmath $\jpsi$ and $\psip$ Hadronic Cross-section Line Shapes}

\begin{figure*}
\includegraphics[width=8cm]{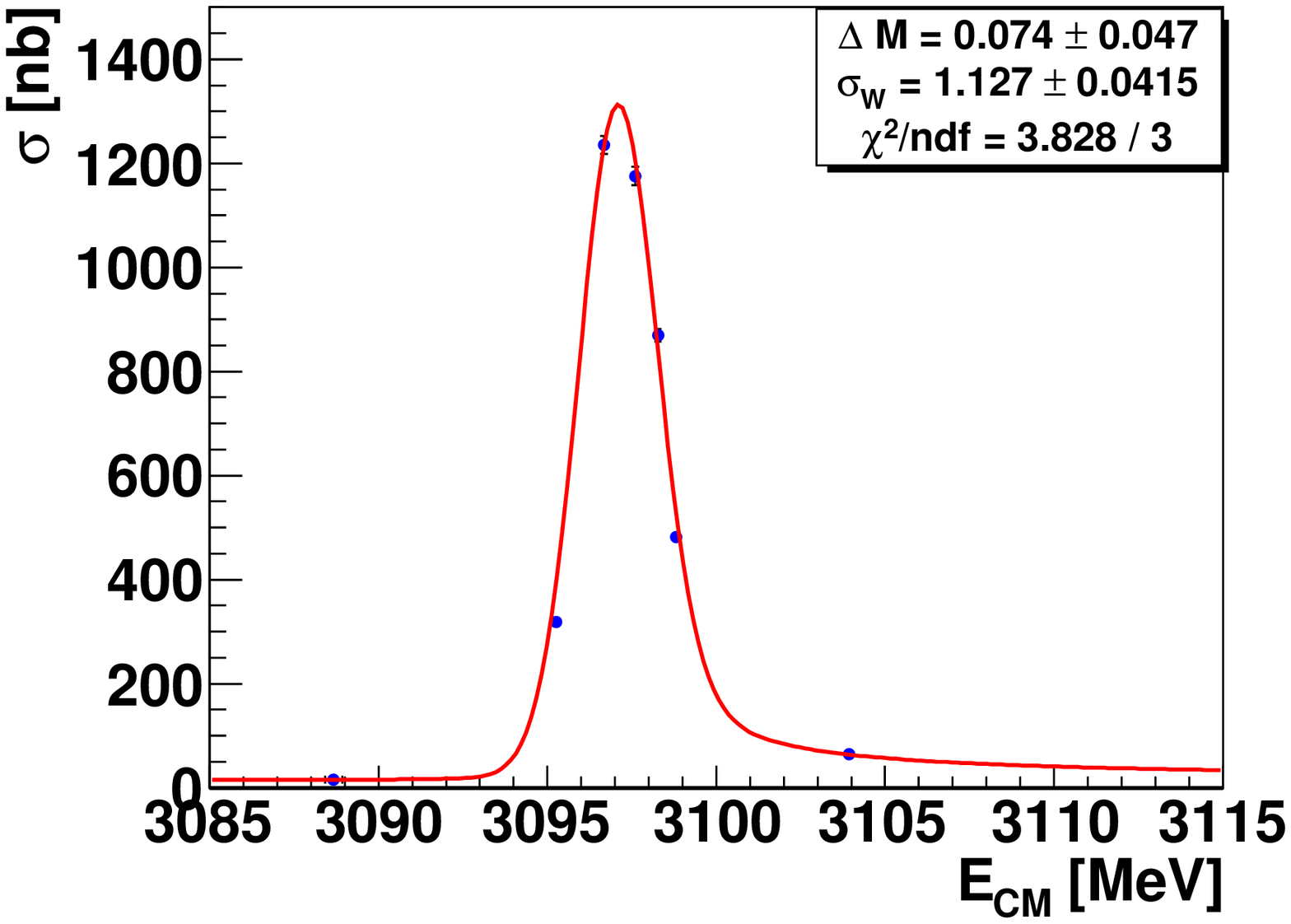}
\includegraphics[width=8cm]{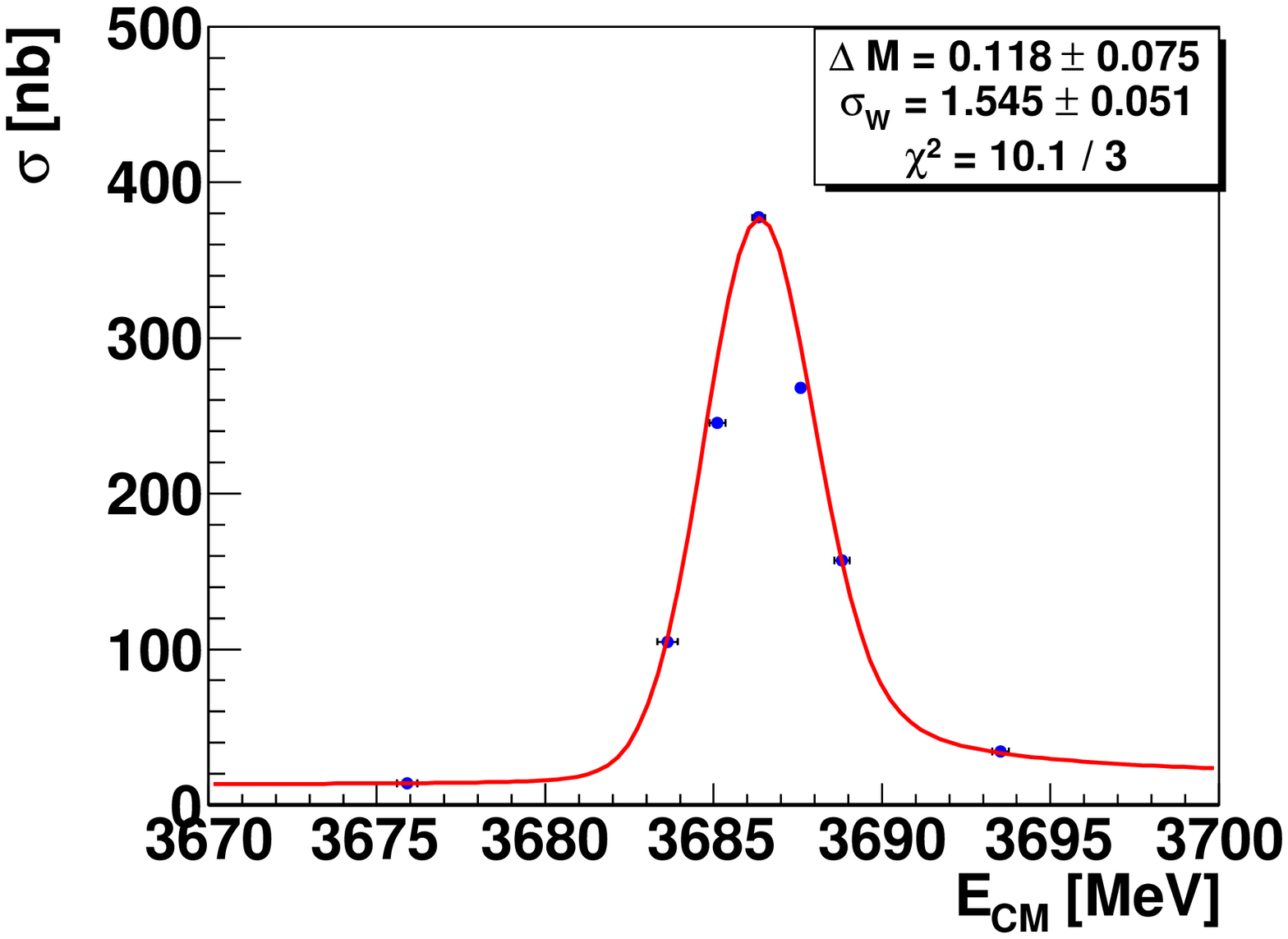}
\caption{\label{fig:fitGG}Fits of the $\jpsi$ (left) and $\psip$ (right) hadronic cross-sections.}
\end{figure*}

The number of hadronic events $N^{h}$ is fitted to the number of
expected hadronic events
\begin{equation}\label{eq:nhad}
N^{exp} = \sigma_{had} \cdot \mathcal{L},
\end{equation}
where $\sigma_{had}$ is the cross section of $\epl \emi \to hadrons$,
which depends on $E_{\rm{CM}}$, and the energy spread, $\delta_w$,
\begin{equation}\label{eq:sigmaHad}
\sigma_{had}(E_{\rm{CM}}, \delta_w) = \sigma_{bg}\cdot \left ( \frac{M}{E_{\rm{CM}}} \right )^{2} + 
                          \epsilon_{had} \cdot \sigma_{res}(E_{\rm{CM}},M,\delta_w).
\end{equation}
Here, $M$ is the resonance mass, and the resonance cross section,
$\sigma_{res}$, is obtained from the hadronic cross section,
$\sigma_0$, described in Ref.~\cite{Tod2009}, taking into account
radiative corrections. The hadronic cross section is convoluted with a
Gaussian with a width equal to the beam energy spread:
\begin{equation}\label{eq:sigRez}
\sigma_{rez} = \int_{-\infty}^{+\infty} 
\frac{e^{-\frac{1}{2}\left ( \frac{E_{\rm{CM}}-E_{\rm{CM}}'}{\delta_w} \right )^2}}{\sqrt{2\pi}\delta_w} \sigma_0(E_{\rm{CM}}',M) dE_{\rm{CM}}'.
\end{equation}
The background cross section, $\sigma_{bg}$, reconstruction
efficiency, $\epsilon_{had}$, $M$, and $\delta_w$, are free 
parameters, obtained from minimizing
\begin{equation}
\chi^2 = \sum_{i=1}^{N} \frac{(N^{h}_i - \sigma_{had}^i \mathcal{L}_i )^2}{ N^{h}_i ( 1 +  N^{h}_i (\Delta \mathcal{L}_i/\mathcal{L}_i)^2 ) },
\end{equation}
where $\Delta \mathcal{L}_i / \mathcal{L}_i$, $N^{h}_i$,
$\sigma_{had}^i$ and $\mathcal{L}_i$ are the relative luminosity
error, the number of hadron events, the cross section of $\epl \emi
\to hadrons$, and the luminosity at scan point $i$, respectively.
Figure~\ref{fig:fitGG} shows the number of hadronic events from the
$\jpsi$ (left) and $\psip$ (right) regions, fitted to the hadronic
cross-section line shapes given by Eq.~(\ref{eq:sigRez}).  The mass
difference with respect to nominal resonance mass, $\Delta M = M^{fit}
- M^{PDG}$, and the energy spread from the $\jpsi$ and $\psip$ fits
are given in Table~\ref{tab:fits}, where the first error is
statistical and the second is systematic.  The values for $\delta_w$
in Table~\ref{tab:fits} agree well with those in
Table~\ref{tab:wSpreadSP} obtained from the BEMS.

The systematic errors are determined by applying different selection
criteria on the number of hadronic events, and using the Bhabha
luminosity instead of the two-gamma luminosity. In addition,
systematic errors from fitting resonance line-shapes when background
is allowed to interfere (Ref.~\cite{F.A.Berends}) are taken into
account. The systematic error associated with determining scan point
energies from the EC data, estimated by comparing calibration lines
and pulsing lines in the BEMS system, is negligible compared to
statistical errors on CM energies.

We extrapolate the results from the fits of the $\jpsi$ and $\psip$
line shapes to the $\tau$-mass region in order to obtain the energy
correction to the $\tau$-mass.  The systematic error associated with
the energy scale is estimated by extrapolating under two assumptions:
the first one is that the correction has a linear dependence on the
energy and the second one assumes a constant shift.  The linear fit
between data points from Table~\ref{tab:fits} gives the correction to
the $\tau$-mass of $\Delta m_{\tau} = (0.054\pm0.030$)~MeV/$c^2$; the constant
shift gives the correction of $\Delta m_{\tau} = (0.043\pm0.020$)~MeV/$c^2$,
where both errors are statistical. The difference between these two
methods, 0.011~MeV/$c^2$, is taken as the systematic uncertainty related
to the energy determination.  The difference of 0.005~MeV from taking into
account background interference when fitting the $\jpsi$ and $\psip$
line shapes is taken as an additional systematic uncertainty.  Overall,
the systematic error in the energy determination is taken to be 0.012~MeV.
The difference
\begin{equation}\label{eq:tauCor}
 \Delta m_{\tau} = 0.054 \pm 0.030 (stat) \pm 0.012 (sys)\; \rm{MeV}/c^2,
\end{equation}
will be taken into consideration in the measured $\tau$-mass value.

\begin{table}[tb]
\begin{center}
  \caption{Fit parameters from the $\jpsi$ and
    $\psip$ fits, where the first error is statistical and the second
    is systematic. $\Delta M$ is the difference between fitted $\psi$
    mass and the normal value from PDG. All units are in MeV/$c^2$.}
\begin{tabular} {lccc}
\hline \hline
 Scan & $\Delta M$  & $\delta_w$   \\ 
\hline
$\jpsi$  & 0.074$\pm$0.047$\pm$0.043  & 1.127$\pm$0.042$\pm$0.050  \\
$\psip$  & 0.118$\pm$0.076$\pm$0.021  & 1.545$\pm$0.051$\pm$0.069  \\
\hline\hline
\end{tabular}
\label{tab:fits}
\end{center}
\end{table}

%%%%%%%%%%%%%%%%%%%%%%%%%%%%%%%%%%%%%%%%%%%%%%%%%%%%%%%%%%%%%%%%%%%%%%%%%%%

\subsection{$\tau$ Mass Measurement}
\subsubsection{Comparison of the Data and MC Samples}

\begin{table*}[tb]
\begin{center}
  \caption{A comparison of the numbers of
events by final state to those from the $\tau$ pair inclusive MC
sample. The MC sample has been normalized to the data according to
the luminosity at each point, and the numbers of normalized MC events
have been multiplied by the ratio of the overall efficiencies for
identifying $\tau$ pair events for data and MC simulation.}
\begin{tabular} {lccccccccccc}
\hline \hline
{final state} & \multicolumn{2}{c} 1 & \multicolumn{2}{c} 2
  & \multicolumn{2}{c} 3 & \multicolumn{2}{c} 4 & \multicolumn{2}{c} {total} \\
         & Data & MC & Data & MC & Data & MC & Data & MC & Data & MC\\
 \hline
  $ee$     & 0  & 0  & 4  &3.7 &13  &12.2 &84 &76.1 &101 &92.0  \\
  $e\mu$   & 0  & 0  & 8  &9.1 &35  &31.4 &168&192.6&211 &233.1 \\
  $e\pi$   & 0  & 0  & 8  &8.6 &33  &29.7 &202&184.4&243 &222.6 \\
  $eK$     & 0  & 0  & 0  &0.5 &2   &1.8  &16 &16.9 &18  &19.3  \\
  $\mu\mu$ & 0  & 0  & 2  &2.9 &8   &9.2  &49 &56.3 &59  &68.4  \\
  $\mu\pi$ & 0  & 0  & 4  &3.9 &11  &14.1 &89 &86.7 &104 &104.7 \\
  $\mu K$  & 0  & 0  & 0  &0.2 &3   &0.8  &7  &9.0  &10  &10.1  \\
  $\pi\pi$ & 0  & 0  & 1  &2.0 &5   &7.7  &57 &54.0 &63  &63.8  \\
  $\pi K$  & 0  & 0  & 1  &0.3 &0   &0.8  &10 &8.2  &11  &9.3   \\
  $KK$     & 0  & 0  & 0  &0.0 &1   &0.1  &1  &0.3  &2   &0.4   \\
  $e\rho$  & 0  & 0  & 3  &6.1 &19  &20.6 &142&132.0&164 &158.7 \\
  $\mu\rho$& 0  & 0  & 8  &3.3 &8  &11.8 &52 &63.3 &68  &78.5  \\
  $\pi\rho$& 0  & 0  & 5  &3.4 &15  &10.8 &97 &96.0 &117 &110.2 \\
  Total    & 0  & 0  & 44 &44.2&153 &151.2&974&975.7&1171&1171.0\\
\hline\hline
\end{tabular}
\label{numttevtdata}
\end{center}
\end{table*}

\begin{figure*}
\includegraphics[width=7cm]{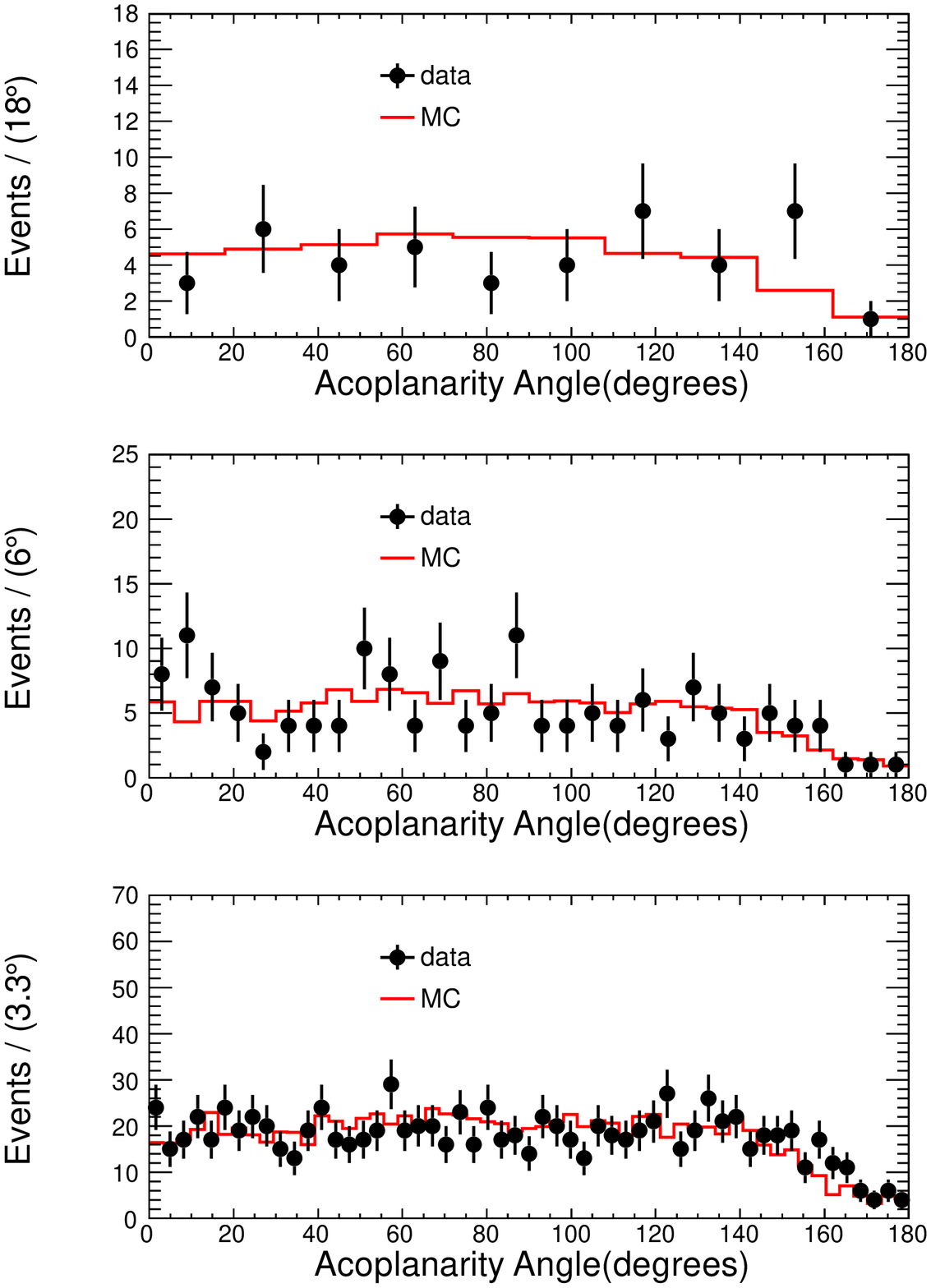}
\includegraphics[width=7cm]{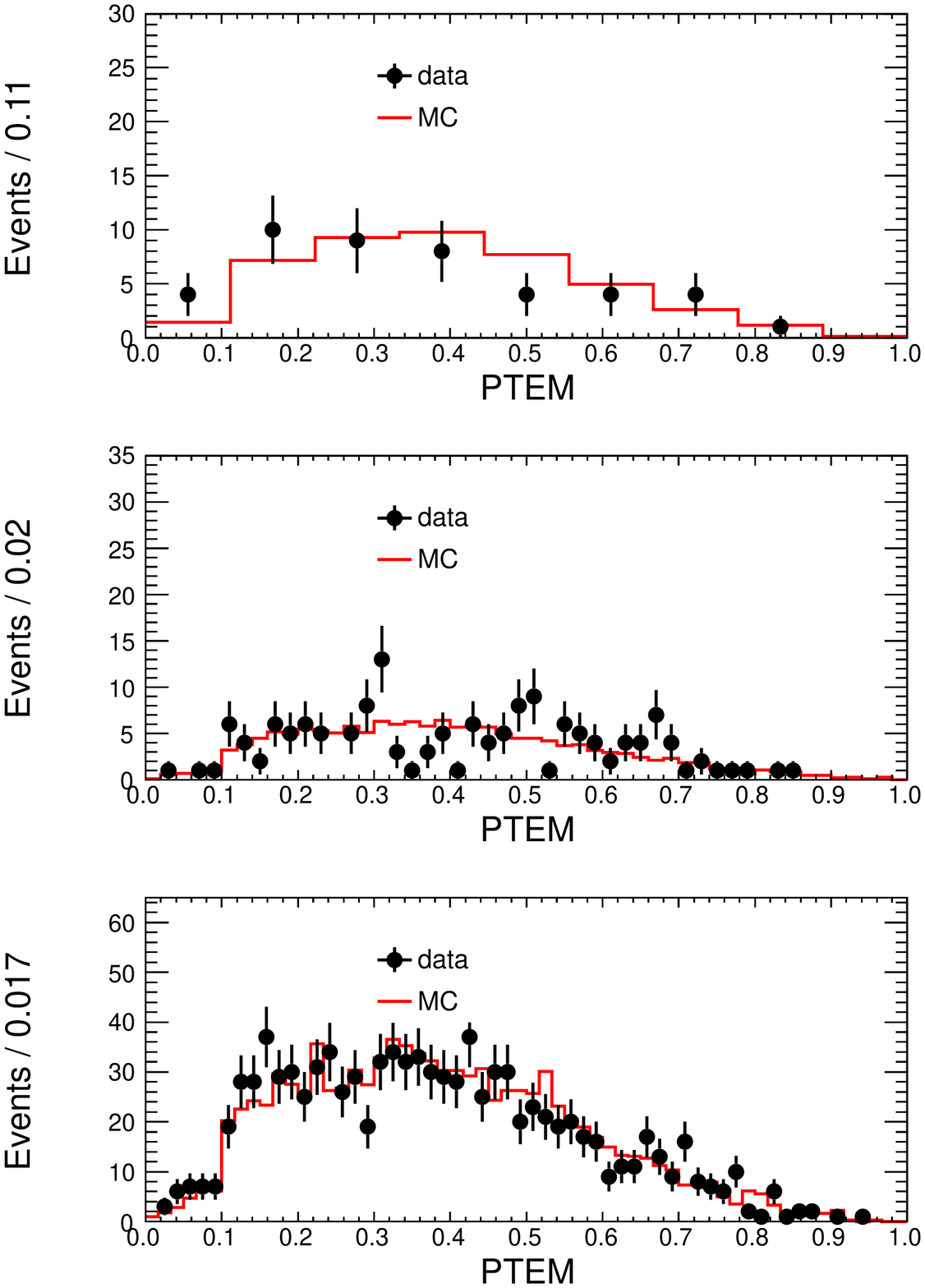}
\caption{\label{acopandptem} (left) The distribution in acoplanarity
  angle between two charged tracks and (right) the distribution in
  PTEM. Dots with error bars are data and the histogram is $\tau$ pair
  inclusive MC. The upper two plots are from the second scan point,
  the middle two are from the third scan point, and the lower two are
  from the fourth scan point.}
\end{figure*}

The comparison between the number of final $\tau$ pair candidate
events from data and from the $\tau$ pair inclusive MC samples are
listed in Table~\ref{numttevtdata} ordered by final state and scan point,
where the indexes in the first row, from 1 to 4 represent the
index of the scan points. The $\tau$ pair inclusive MC sample has been
normalized to the data according to the luminosity at each point, and
the numbers of normalized MC events have been multiplied by the ratio
of the overall efficiencies for identifying $\tau$ pair events for
data and MC simulation, which is fitted from the data set in the
following section.

The comparison of some distributions between data and the $\tau$ pair
inclusive MC samples are shown in Fig.~\ref{acopandptem}.
These comparisons and those in Table~\ref{numttevtdata}
indicate that data and MC samples agree well with each other.

The selected $\tau$ pair candidate events are used for the
measurement of the mass of the $\tau$ lepton and the corresponding
$\tau$ pair inclusive MC samples are used to obtain the selection
efficiency for different decay channels.

%%%%%fah down to here
\subsubsection{Maximum Likelihood Fit to The Data}
%\label{fitting}
\begin{figure*}
\includegraphics[width=6cm]{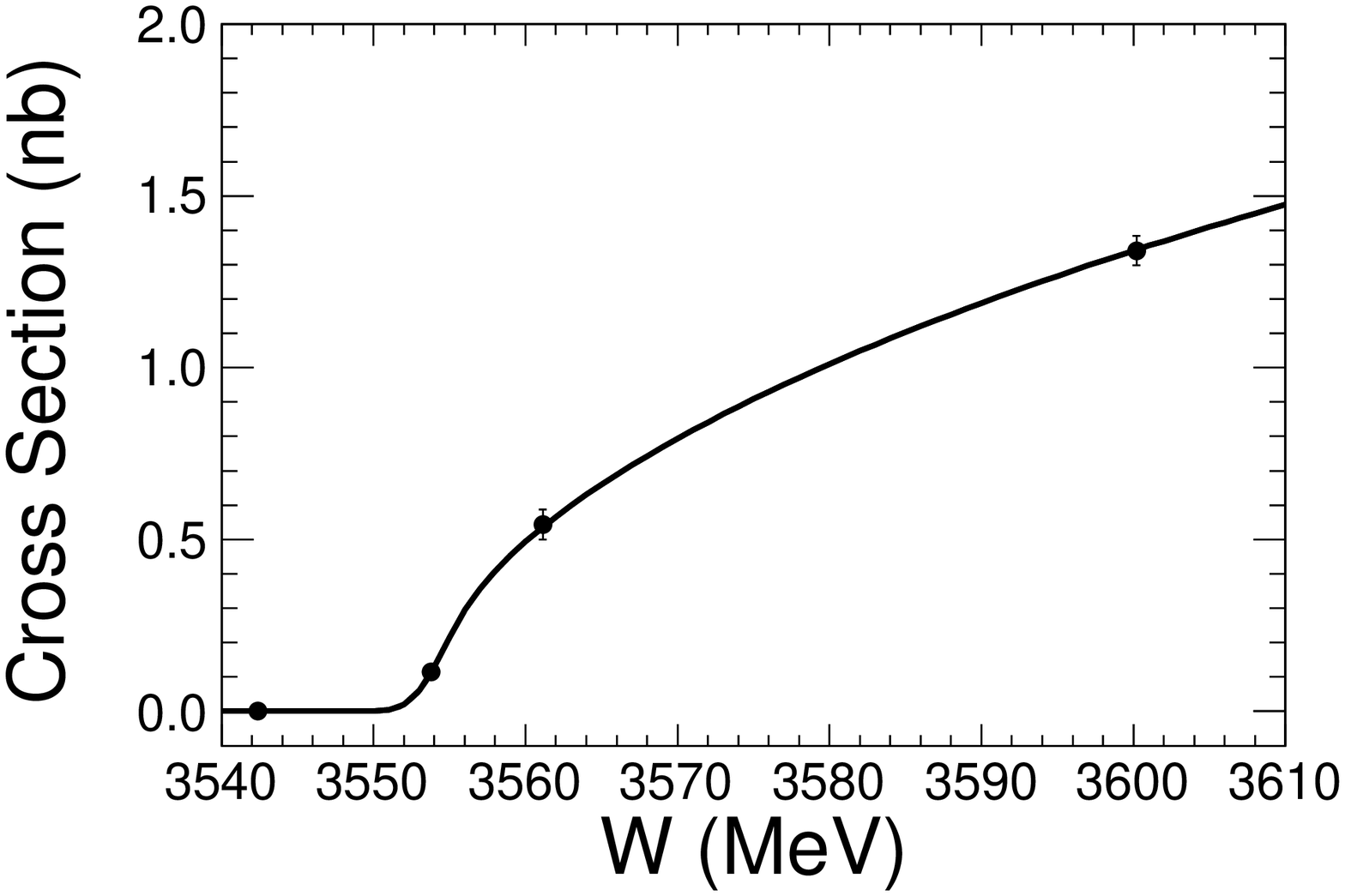}
\includegraphics[width=6cm]{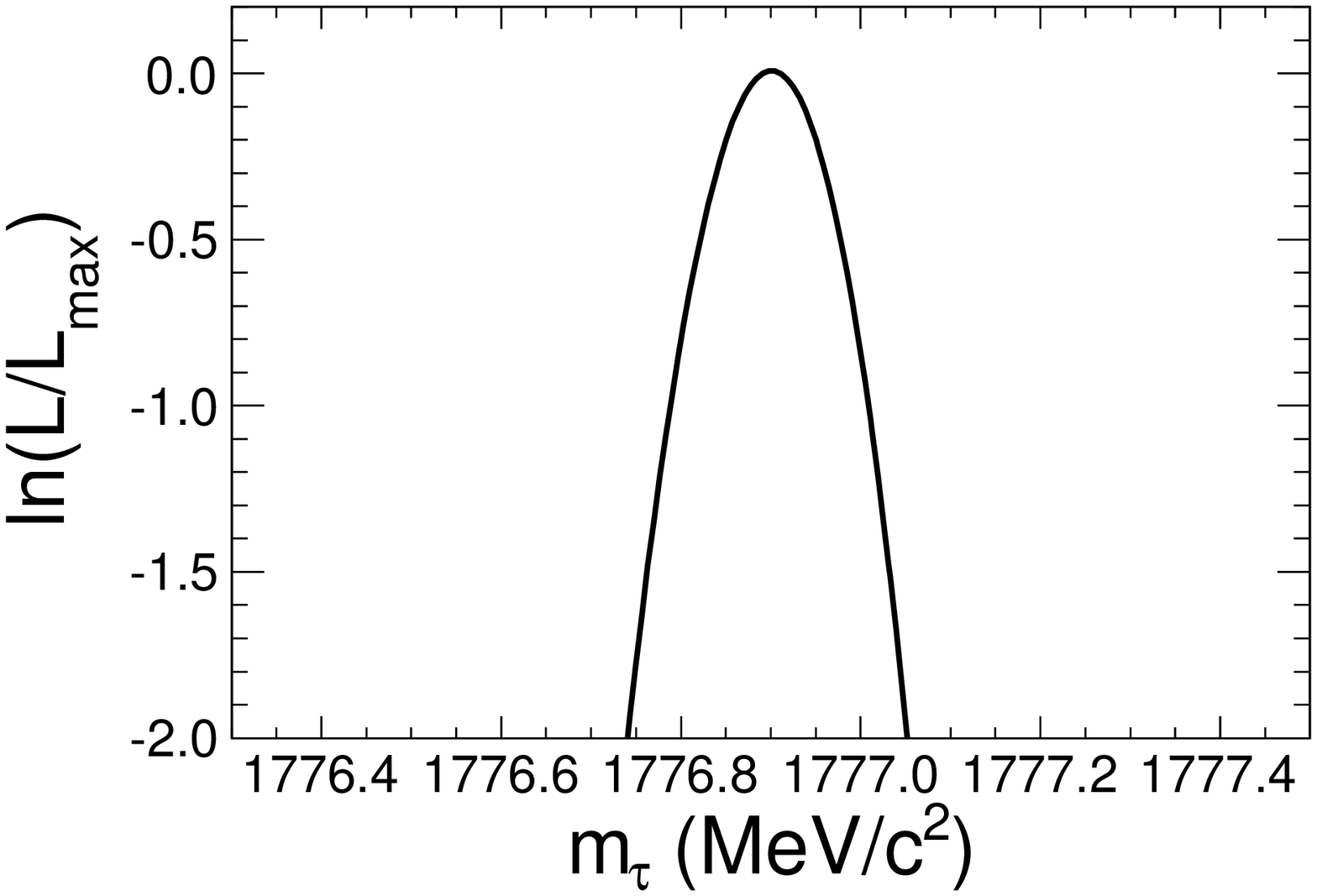}
\caption{\label{fittau} (left) The CM energy dependence of the $\tau$ pair cross
section resulting from the likelihood fit (curve), compared to the
data (Poisson errors), and (right) the dependence of the logarithm of
the likelihood function on $m_{\tau}$, with the efficiency and
background parameters fixed at their most likely values.}
\end{figure*}

The mass of the $\tau$ lepton is obtained from a
maximum likelihood fit to the CM energy dependence of the $\tau$
pair production cross section. The likelihood function is
constructed from Poisson distributions, one at each of the four scan
points, and takes the form~\cite{bes2prd}

\begin{eqnarray}
\label{likelihood} L(m_{\tau}, {\cal R}_{Data/MC}, \sigma_{B}) =
\prod^{4}_{i=1}\frac{\mu^{N_i}_ie^{-\mu_i}}{N_i!},
\end{eqnarray}
where $N_i$ is the number of observed $\tau$ pair events at scan
point $i$; $\mu_i$ is the expected number of events and calculated
by

\begin{eqnarray}
\label{expectednum} \mu_i=[{\cal
R}_{Data/MC}\times\epsilon_i\times\sigma(E_{\rm{CM}}^i,m_{\tau})+\sigma_B]\times{\cal
L}_i.
\end{eqnarray}
In Eq.~(\ref{expectednum}), $m_{\tau}$ is the mass of the $\tau$
lepton, and $\RDM$ is the ratio of the overall efficiency for
identifying $\tau$ pair events for data and for MC simulation,
allowing for the difference of the efficiencies between the data and
the corresponding MC sample. $\epsilon_i$ is the efficiency at scan
point $i$, which is given by $\epsilon_i=\sum_i Br_j\epsilon_{ij}$,
where $Br_j$ is the branching fraction for the $j-th$ final state and
$\epsilon_{ij}$ is the detection efficiency for the $j-th$ final state
at the $i-th$ scan point. The efficiencies $\epsilon_i$, determined
directly from the $\tau$ pair inclusive MC sample by applying the same
$\tau$ pair selection criteria, are 0.065, 0.065, 0.069, 0.073 at the
four scan points, respectively.  $\sigma_B$ is an effective background
cross section, and it is assumed constant over the limited range of CM
energy, $E_{\rm{CM}}^i$, covered by the scan. ${\cal L}_i$ is the
integrated luminosity at scan point $i$, and
$\sigma(E_{\rm{CM}}^i,m_{\tau})$ is the corresponding cross section for
$\tau$ pair production which has the form~\cite{bes2prd}

\begin{widetext}
\begin{eqnarray}
\label{tautaucrosssec}
\sigma(E_{\rm{CM}},m_{\tau},\delta_w^{\rm{BEMS}})=\frac{1}{\sqrt{2\pi}\delta_w^{\rm{BEMS}}}\int_{2m_{\tau}}^{\infty}dE_{\rm{CM}}^{\prime}
e^{\frac{-(E_{\rm{CM}}-E_{\rm{CM}}^{\prime})^2}{2(\delta_w^{\rm{BEMS}})^2}}\int_{0}^{1-\frac{4m^2}{E_{\rm{CM}}^{\prime
2}}}dxF(x,E_{\rm{CM}}^{\prime})\frac{\sigma_1(E_{\rm{CM}}^{\prime}\sqrt{1-x},m_{\tau})}{|1-\prod(E_{\rm{CM}})|^2}.
\end{eqnarray}
\end{widetext}

Here, $\delta_w^{\rm{BEMS}}$ is the CM energy spread, determined from the BEMS;
$F(x,E_{\rm{CM}})$ is the initial state radiation factor~\cite{E.A.Kuraev};
$\prod(E_{\rm{CM}})$ is the vacuum polarization factor~\cite{P.Ruiz,
F.A.Berends, G.Rodrigo}; and $\sigma_1(E_{\rm{CM}},m_{\tau})$ is the high
accuracy, improved cross section from Voloshin~\cite{newVoloshin}.
In carrying out the maximum likelihood (ML) fit, $m_{\tau}$, $\RDM$ and $\sigma_B$ are
allowed to vary, subject to the requirement $\sigma_B\geq$0.

To test the procedure, the likelihood fit is performed on the selected
$\tau$ pair inclusive MC data sample. The input $m_{\tau}$ is $1776.90$~MeV/$c^2$,
    while the fitted value of $m_{\tau}$ is found to be
$m_{\tau}=(1776.90\pm0.12$)~MeV/$c^2$; the good agreement between the
input and output values indicates that the fitting procedure is
reliable.

The same ML fit is performed on the selected $\tau$ pair candidate
events.
The fit yields

\begin{eqnarray}
\label{fittedrlt} &m_{\tau}&=1776.91\pm0.12 \mbox{ MeV/$c^2$}, \nonumber\\
                    &{\RDM}&=1.05\pm0.04, \nonumber\\
                  &\sigma_B&=0^{+0.12} \mbox{ pb}.
\end{eqnarray}
The fitted $\sigma_B$ is zero, which indicates the selected
$\tau$ pair candidate data set is very pure.

The quality of the fit is shown explicitly in Fig.~\ref{fittau} (left
plot). The curve corresponds to the cross section given by
Eq.~(\ref{tautaucrosssec}) with $m_{\tau}=1776.91$~MeV/$c^2$; the measured
cross section at scan point $i$ is given by
\begin{eqnarray}
\label{measurednum} \sigma_i=\frac{N_i}{\RDM\epsilon_i{\cal L}_i}.
\end{eqnarray}
The measured cross sections at different scan points are consistent
with the theoretical values.  In Fig.~\ref{fittau} (right plot), the
dependence of $\ln L$ on $m_{\tau}$ is almost symmetric as a
consequence of the large data sample obtained.

\subsubsection{Systematic Error Estimation}
%Since the efficiency parameter $\RDM$ and the effective background
%cross section $\sigma_B$ are floated in the fitting, the uncertainties
%from these two items are included in the statistical errors, and they
%will not be considered here.
\paragraph{Theoretical Accuracy}
The systematic error associated with the theoretical $\tau$ pair
production cross section is estimated by comparing the difference of
the fitted $m_{\tau}$ between two cases; in one case, the old
$\tau$ pair production cross section formulas are used, in the other,
the improved version formulas are used. The uncertainty due to this
effect is at the level of $10^{-3}$~MeV/$c^2$. More details can be
found in reference~\cite{xiaohu}.

\paragraph{Energy Scale}
%\label{enescale}
The $m_{\tau}$ shift, $\Delta \tau_M = (0.054 \pm 0.030 (stat) \pm 0.012 (sys)$)~MeV/$c^2$
(Eq.~\ref{eq:tauCor}) is taken as a systematic error. Combining 
statistical and systematic errors, two boundaries can be established:
$\Delta m_{\tau}^{low} = 0.054 - 0.032 = 0.022$~MeV/$c^2$ and $\Delta
m_{\tau}^{high} = 0.054 + 0.032 = 0.086$~MeV/$c^2$. We take the higher
value to form a negative systematic error and the lower value the positive
systematic error.  The systematic errors on the $m_{\tau}$ from this
source are $\Delta m_{\tau}^{-} = 0.086$~MeV/$c^2$ and $\Delta m_{\tau}^{+} =
0.022$~MeV/$c^2$.

\paragraph{Energy Spread}
From Table~\ref{tab:wSpreadSP}, $\delta_w^{\rm{BEMS}}$ at the
$\tau$ scan energy points is determined from the BEMS to be
($1.469\pm0.064$)~MeV. If we assume quadratic dependence of $\delta_w$
on energy, we can also extrapolate the $J/\psi$ and $\psi'$ energy
spreads to the $\tau$ region, which yields $\delta_w = (1.471\pm0.040$)~MeV.
The difference of energy spreads obtained from these two methods
is taken as a systematic uncertainty. The largest contribution to the
energy spread uncertainty comes from interference effects. Including
interference, the difference between the extrapolated value and the
BEMS measurement is 0.056~MeV, and the overall systematic error is taken
as 0.057~MeV. The final energy spread at the $\tau$ scan energy points
is ($1.469\pm0.064\pm0.057$)~MeV. The uncertainty of $m_{\tau}$
from this item is estimated by refitting the data when the
energy spread is set at its $\pm1\sigma$ values, and the shifted value
of the fitted $m_{\tau}$, $\pm0.016$~MeV/$c^2$, is taken as the
systematic error. Table~\ref{sys-enespread} lists the fitted results
with different energy spread values.

\begin{table}[tb]
\begin{center}
  \caption{The $\tau$ mass determined from
fits with different energy spreads.}
\begin{tabular} {lc}
\hline \hline
$\delta_w^{\rm{BEMS}}$ (MeV) & $\tau$ mass (MeV/$c^2$) \\
\hline
  1.383 & 1776.891$_{-0.117}^{+0.111}$\\
  1.469 & 1776.906$_{-0.120}^{+0.116}$\\
  1.553 & 1776.919$_{-0.126}^{+0.119}$\\
\hline\hline
\end{tabular}
\label{sys-enespread}
\end{center}
\end{table}

\paragraph{Luminosity}
Both the Bhabha and the two-gamma luminosities are used in fitting the
$\tau$ mass, and the difference of fitted $\tau$ masses is taken as
the systematic error due to uncertainty in the luminosity
determination. The difference is $0.001$~MeV/$c^2$.
%, which can be found in Table~\ref{sys-lum}.

The $\tau$ mass shift (Eq.~\ref{eq:tauCor}) is 0.054~MeV/$c^2$ when
determined with two-gamma luminosities. If instead, Bhabha
luminosities are used, the mass shift is 0.059~MeV/$c^2$, and the
difference, 0.005~MeV/$c^2$, is also taken as a systematical error due
to the luminosity.  The total systematical uncertainty from luminosity
determination is 0.006~MeV/$c^2$.

\paragraph{Number of Good Photons}
It is required that there are no extra good photons in our final
states. Bhabha events are selected as a control sample to study the
efficiency difference between data and MC of this requirement. The
efficiency for data is (79.17$\pm$0.06)\%, and the efficiency for the
MC simulation is (79.01$\pm$0.14)\%, where the errors are statistical.
Correcting the number of observed events from data for the efficiency
difference, we refit the $\tau$ mass, and the change of $\tau$ mass is
0.002~MeV/$c^2$, which is taken as the systematic uncertainty for this
requirement.

\paragraph{PTEM and Acoplanarity Angle Requirements}
The nominal selection criteria on PTEM and Acoplanarity Angle,
which are described in Section~\ref{subsubsec-selcut}, are
determined based on the first scan point data.
The $\tau$ mass is refitted using an alternative selection, where
the requirements on PTEM and Acoplanarity Angle
have been optimized based on MC simulation, and the change of the
fitted $\tau$ mass from the nominal value, 0.05~MeV/$c^2$, is taken
as the systematic error.

\paragraph{Mis-ID Efficiency}

To determine the systematic error from misidentification between
channels, two fits are done. In the first (nominal) fit, we use 
the particle ID efficiencies and misidentification (mis-ID) rates as obtained
from $\tau$ pair inclusive MC samples. For the second fit, we extract
PID efficiencies and mis-ID rates from selected data control samples of
radiative Bhabha events, $J/\psi \to \rho \pi$, and cosmic ray events, 
correct the selection efficiencies of the different $\tau$ pair final
states and propagate these changes to the event selection efficiencies
$\epsilon_i$. We then refit our data with these modified efficiencies.
The difference between the fitted $\tau$ mass from these two fits, 
0.048~MeV/$c^2$, is taken as the systematic error due to misidentification
between different channels.

\paragraph{Background Shape}
In this analysis, the background cross section $\sigma_{B}$ is assumed
to be constant for different $\tau$ scan points. The background cross
sections have also been estimated at the last three scan points by
applying their selection criteria on the first scan point data, where
the $\tau$ pair production is zero. After fixing $\sigma_B$ to these
values, the fitted $\tau$ mass becomes:
\begin{eqnarray}
\label{fixbkgfittedrlt} &m_{\tau}&=(1776.87\pm0.12) \mbox{ MeV/$c^2$},  
\end{eqnarray}
The fitted $\tau$ mass changed by 0.04~MeV/$c^2$ compared to the
nominal result.

\paragraph{Fitted Efficiency Parameter}
The systematic uncertainties associated with the fitted efficiency parameter 
are obtained by setting ${\cal R}_{Data/MC}$ at its $\pm 1\sigma$ value and maximizing
the likelihood with respect to $m_{\tau}$ with $\sigma_B$ = 0. This method 
yields changes in the fitted $\tau$ mass of $\Delta m_{\tau} = ^{+0.038}_{-0.034}$~MeV, 
which is taken as the systematic uncertainty.

\paragraph{Total Systematic Error}

The systematic error sources and their contributions are summarized in
Table~\ref{total-err}. We assume that all systematical uncertainties are
independent and add them in quadrature to obtain the total systematical
uncertainty for $\tau$ mass measurement, which is $^{+0.10}_{-0.13}$ MeV/$c^2$.

\begin{table}[tb]
\begin{center}
  \caption{Summary of the  $\tau$ mass systematic errors.}
\begin{tabular} {lc}
\hline \hline
Source & $\Delta m_{\tau}$ (MeV/$c^2$) \\
\hline
  Theoretical accuracy & 0.010 \\
  Energy scale & $^{+0.022}_{-0.086}$  \\
  Energy spread & 0.016 \\
  Luminosity & 0.006 \\
  Cut on number of good photons & 0.002 \\
  Cuts on PTEM and acoplanarity angle & 0.05 \\
  %Selection requirement & 0.008 \\
  mis-ID efficiency    & 0.048 \\
  Background shape & 0.04 \\
  Fitted efficiency parameter & $^{+0.038}_{-0.034}$\\
  \hline
  Total & $^{+0.094}_{-0.124}$\\\hline\hline
\end{tabular}
\label{total-err}
\end{center}
\end{table}

%%%%%%%%%%%%%%%%%%%%%%%%%%%%%%%%%%%%%%%%%%%%%%%%%%%%%%%%%%%%%%%%%%%%%%%%%%%

\section{Results}

\begin{figure}
\includegraphics[width=8cm]{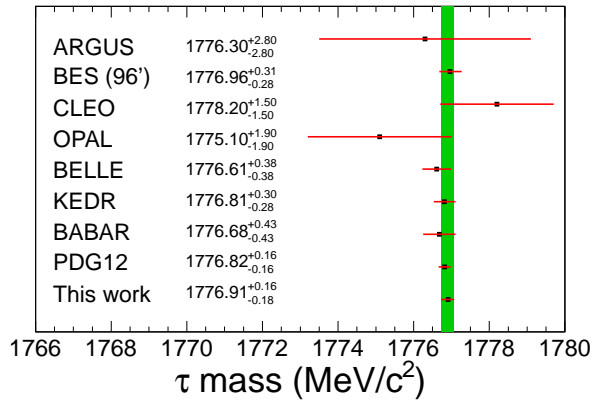}
\caption{\label{comparedtaumass} Comparison of measured $\tau$ mass
  from this paper with those from the PDG. The green band corresponds
      to the 1~$\sigma$ limit of the measurement of this paper}
\end{figure}

By a maximum likelihood fit to the $\tau$ pair cross section data
near threshold, the mass of the $\tau$ lepton has been measured as
\begin{eqnarray}
\label{finaltaumass} m_{\tau} = (1776.91\pm0.12^{+0.10}_{-0.13}) \mbox{ MeV/$c^2$}.
\end{eqnarray}
Figure~\ref{comparedtaumass} shows the comparison of measured $\tau$
mass in this paper with values from the PDG~\cite{pdg}; our result is
consistent with all of them, but with the smallest uncertainty.

Using our $\tau$ mass value, together with the values of $B(\tau\rightarrow e\nu\bar{\nu})$ and 
$\tau_{\tau}$ from the PDG~\cite{pdg}, we can calculate $g_{\tau}$  
through Eq.~\ref{Eq.calgtau}: 
\begin{eqnarray}
\label{Eq.gtau} g_{\tau}=(1.1650\pm0.0034)\times 10^{-5} ~\rm{GeV}^{-2},  
\end{eqnarray}
which can be used to test the SM.

Similarly, inserting our $\tau$ mass value into Eq.~\ref{Eq.universaltest} ,
together with the values of $\tau_{\mu}$, $\tau_{\tau}$, $m_{\mu}$,
$m_{\tau}$, $B(\tau\rightarrow e\nu\bar{\nu})$ and $B(\mu\rightarrow e\nu\bar{\nu})$
from the PDG~\cite{pdg} and using the values of $F_{W}$ (-0.0003) and
$F_{\gamma}$ (0.0001) calculated from
reference~\cite{LepUnivCor}, the ratio of squared coupling constants is
determined to be: 
\begin{eqnarray}
\label{Eq.lepuniversal} \left(\frac{g_{\tau}}{g_{\mu}}\right)^{2}=1.0016\pm0.0042,  
\end{eqnarray}
so that this test of lepton universality is satisfied at the 0.4 standard deviation
level. The level of precision is compatible with previous
determinations, which used the PDG average for $m_{\tau}$~\cite{pich}.

\section{Acknowledgments}
    The BESIII collaboration thanks the staff of BEPCII and the computing center for their strong support. This work is supported in part by the Ministry of Science and Technology of China under Contract No. 2009CB825200; Joint Funds of the National Natural Science Foundation of China under Contracts Nos. 11079008, 11179007, U1332201; National Natural Science Foundation of China (NSFC) under Contracts Nos. 11375206, 10625524, 10821063, 10825524, 10835001, 10935007, 11125525, 11235011, 11179020; the Chinese Academy of Sciences (CAS) Large-Scale Scientific Facility Program; CAS under Contracts Nos. KJCX2-YW-N29, KJCX2-YW-N45; 100 Talents Program of CAS; German Research Foundation DFG under Contract No. Collaborative Research Center CRC-1044; Istituto Nazionale di Fisica Nucleare, Italy; Ministry of Development of Turkey under Contract No. DPT2006K-120470; U. S. Department of Energy under Contracts Nos. DE-FG02-04ER41291, DE-FG02-05ER41374, DE-FG02-94ER40823, DESC0010118; U.S. National Science Foundation; University of Groningen (RuG) and the Helmholtzzentrum fuer Schwerionenforschung GmbH (GSI), Darmstadt; WCU Program of National Research Foundation of Korea under Contract No. R32-2008-000-10155-0.


\begin{thebibliography}{99}

\bibitem{LepUnivCor} W.J. Marciano and A. Sirlin, Phys. Rev. Lett. \textbf{61}
, 1815 (1988).

\bibitem{bes1} J.Z. Bai {\em et al.}, (BES Collaboration), Phys. Rev. Lett. {\bf 69}, 3021
  (1992).

\bibitem{bes2prd} J.Z. Bai {\em et al.}, (BES Collaboration), Phys. Rev. D {\bf 53}, 20 (1996).
\bibitem{belle} K. Belous \textit{et al.}, (Belle Collaboration), Phys. Rev. Lett., \textbf{99}, 011801 (2007).
\bibitem{kedr} V.V. Anashin {\em et al.}, (KEDR Collaboration), JETP Letters {\bf 85}, 347 (2007).
\bibitem{BABAR} B. Aubert {\em et al.}, (BABAR Collaboration), Phys. Rev. D {\bf 80}, 092005 (2009) 
\bibitem{pdg} J. Beringer {\em et al.}, (Particle Data Group), Phys. Rev. D {\bf 86}, 010001 (2012)
and 2013 partial update for the 2014 edition.

\bibitem{mntau} R. Barate \textit{et al.}, Eur. Phys. J. C \textbf{2}, 395 (1998).
\bibitem{argus} H. Albrecht \textit{et al.}, (ARGUS Collaboration), Phys. Lett. B \textbf{292}, 221 (1992). 
\bibitem{opal} G. Abbiendi \textit{et al.}, (OPAL Collaboration), Phys. Lett. B \textbf{492}, 23 (2000). 
\bibitem{delco} W. Bacino {\em et al.}, (DELCO Collab.), Phys. Rev. Lett. {\bf 41}, 13 (1978).
\bibitem{bems} E.V. Abakumova {\em et al.}, Nucl. Instrum. Methods Phys. Res., Sect. A {\bf 659}, 21 (2011).
\bibitem{optstd3}
M.N.~Achasov {\em et al.}, Chinese Phys. C \textbf{36}, 573 (2012).
\bibitem{bes3} M. Ablikim {\em et al.}, Nucl. Instrum. Methods Phys. Res., Sect. A \textbf{614}, 345 (2010).
\bibitem{Rullhusen}
P.~Rullhusen, X.~Artru, P.~Dhez, Novel Radiation Sources
Using Relativistic Electrons (World Scientific Publishing, Singapore, 1998).
\bibitem{Landau}
L.D.~Landau, E.M.~Lifshitz, Relativistic Quantum Mechanics~(Pergamon, New York, 1971).

\bibitem{babayaga} Giovanni Balossini {\em et al.}, Nucl. Phys. B \textbf{758}, 227 (2006).

\bibitem{geant4} S. Agostinelli {\em et al.} (GEANT4 Collaboration),
  Nucl. Instrum. Methods Phys. Res., Sect. A {\bf 506}, 250 (2003).

\bibitem{boss} Z. Y. Deng {\em et al.}, HEP\&NP {\bf 30}, 371 (2006).

\bibitem{kkmc} S.~Jadach, B.F.L. Ward and Z.~Was, Comput. Phys. Commun. {\bf 130}, 260 (2000);
               Phys. Rev. D {\bf 63}, 113009 (2001).
\bibitem{evtgen} D.J.~Lange {\em et al.}, Nucl. Instrum. Methods Phys. Res., Sect. A {\bf 462}, 152 (2001).
\bibitem{lund} J.C. Chen {\em et al.}, Phys. Rev. D {\bf 62}, 034003 (2000).

%\bibitem{Tod2009}
%K.Yu. Todyshev, \newblock {The application Breit-Wigner form with radiative corrections to the
%      resonance fitting}.
%      \newblock (2009).
\bibitem{Tod2009}
K.Yu. Todyshev, arXiv:0902.4100
      \newblock (2009).
\bibitem{F.A.Berends} F.A. Berends, K.J.F. Gaemers and R. Gastmans,
Nucl. Phys. \textbf{B57}, 381 (1973); F.A. Berends and G.J. Komen, Phys.
Lett. \textbf{63B}, 432 (1976).

\bibitem{E.A.Kuraev} E.A. Kuraev {\em et al.}, Sov.~J.~Nucl.~Phys.
\textbf{41}, 466 (1985); O. Nicrosini and L. Trentadue, Phys. Lett. B \textbf{196}, 551 (1987); 
F.A. Berends, G. Burgers and W.L. Neerven, Nucl. Phys. \textbf{B297}, 429 (1988);
Nucl. Phys. \textbf{B304}, 921 (1988).

\bibitem{P.Ruiz} P. Ruiz-Femen\'{\i}a and A. Pich, Phys. Rev. D \textbf{64}, 053001 (2001).
\bibitem{G.Rodrigo} G. Rodrigo, A. Pich and A. Santamaria, Phys. Lett. B \textbf{424}, 367 (1998).
\bibitem{newVoloshin} M.B. Voloshin, Phys Lett. B \textbf{556}, 153 (2003).
\bibitem{xiaohu} X.H. Mo, Nucl. Phys. Proc. Suppl. \textbf{169}, 132 (2007);
Y.K. Wang {\em et al.}, HEP\&NP {\bf 31}, 325 (2007).

\bibitem{pich} A. Pich, Acta Phys. Polon. B {\bf 38}, 3449 (2007).

\end{thebibliography}
\end{document}